\documentclass[twocolumn,showpacs,prb]{revtex4}
\usepackage{amsfonts}
\usepackage{amsmath}
\usepackage{amssymb}
\usepackage{graphicx}

\begin{document}

\title{Rashba spin splitting in biased semiconductor quantum wells}
\author{W. Yang}
\author{Kai Chang}
\altaffiliation[Author to whom correspondence should be addressed. Electronic address: ]{kchang@red.semi.ac.cn}
\affiliation{NLSM, Institute of Semiconductors, Chinese Academy of Sciences, P. O. Box
912, Beijing 100083, China}

\begin{abstract}
Rashba spin splitting (RSS) in biased semiconductor quantum wells is
investigated theoretically based on the eight-band envelope function model.
We find that at large wave vectors, RSS is both
nonmonotonic and anisotropic as a function of in-plane wave vector, in contrast to the widely used linear and isotropic model.
We derive an analytical expression for RSS,
which can correctly reproduce such nonmonotonic behavior at large wave vectors. We also
investigate numerically the dependence of RSS on the various band parameters and find that
RSS increases with decreasing band gap and subband index, increasing valence band offset, external electric field, and well width.
Our analytical expression for RSS provides a satisfactory explanation to all these features.
\end{abstract}

\maketitle

Rashba spin splitting (RSS) due to structure inversion asymmetry has
attracted intense interest recently,\cite{Wolf} due to its potential
application in spintronic devices.\cite{Datta,Loss,Nitta} The origin of RSS,
however, has been a controversial issue for over 30 years.\cite%
{Ohkawa,Darr,Vasko,Lassnig,Sobkowicz,Shaper}. Recently, it was shown that RSS in
the conduction band (CB) was determined by the total electric field in the
valence band (VB).\cite{book} Approximate analytical expressions for RSS
were also derived\cite{Bassani,Pfeffer2}. These analytical expressions are
helpful for understanding RSS. However, they only give rough estimates of
the magnitude of RSS. Most of the features of the numerical results for RSS,%
\cite{Bassani,Pfeffer,Rossler,Lamari,Zhang,Gui,Pfeffer2} e.g., the
nonmonotonic behaviors at large wave vectors\cite{Zhang} and the dependence
on the subband and well width \cite{Bassani} can not be explained through
these expressions. Further, in order to control the magnitude of RSS, it is
important to understand the dependence of RSS on the various band
parameters. However, to the best of our knowledge, this issue has not been
clearly discussed.

In this Rapid Communication, we present a theoretical study of RSS in biased
quantum wells in the framework of eight-band envelope function model. The full 
numerical results show that at large wave vectors, RSS is both
nonmonotonic and anisotropic as a function of in-plane wave vector $\mathbf{k}_{\parallel }$ ,
in contrast to the widely used linear and monotonic model. We derive an analytical expression for RSS, which
can correctly reproduce such nonmonotonic behavior at large wave vectors. Our analytical expression shows that
such behavior comes from the energy dispersion of the subband.
We also investigate numerically the dependence of RSS on the various band parameters and find
that RSS increases with decreasing band gap and
subband index, increasing valence band offset (VBO), external electric
field, and well width. Our analytical expression gives better descriptions to all these dependences
than the previous analytical expressions.\cite{Bassani,Pfeffer2}

In the basis
\begin{align*}
u_{1}& =S\uparrow , \\
u_{2}& =S\downarrow , \\
u_{3}& =\left\vert 3/2,3/2\right\rangle =(X+iY)\uparrow /\sqrt{2}, \\
u_{4}& =\left\vert 3/2,1/2\right\rangle =i/\sqrt{6}\left[ (X+iY)\downarrow
-2Z\uparrow \right] , \\
u_{5}& =\left\vert 3/2,-1/2\right\rangle =1/\sqrt{6}\left[ (X-iY)\uparrow
+2Z\downarrow \right] , \\
u_{6}& =\left\vert 3/2,-3/2\right\rangle =i/\sqrt{2}(X-iY)\downarrow , \\
u_{7}& =\left\vert 1/2,1/2\right\rangle =1/\sqrt{3}\left[ (X+iY)\downarrow
+Z\uparrow \right] , \\
u_{8}& =\left\vert 1/2,-1/2\right\rangle =i/\sqrt{3}\left[ -(X-iY)\uparrow
+Z\downarrow \right] ,
\end{align*}%
for [001] orientated heterostructures, the eight-band Hamiltonian $H=H_{%
\mathbf{k}=0}+H_{\mathbf{k}}+V$, where $V=eFz$ is the external electric
field-induced potential, $H_{\mathbf{k}=0}=$diag$\left\{
E_{c},E_{c},E_{v},E_{v},E_{v},E_{v},E_{v}-\Delta ,E_{v}-\Delta \right\} $ is
the band-edge Hamiltonian, and the upper triangle of $H_{\mathbf{k}}$ reads 
\begin{widetext}
\[
H_{\mathbf{k}}=
\begin{bmatrix}
\mathbf{k}A_{c}\mathbf{k} & 0 & iP_{0}k_{+}/\sqrt{2} & \sqrt{2/3}P_{0}k_{z} &
iP_{0}k_{-}/\sqrt{6} & 0 & iP_{0}k_{z}/\sqrt{3} & P_{0}k_{-}/\sqrt{3}\\
& \mathbf{k}A_{c}\mathbf{k} & 0 & -P_{0}k_{+}/\sqrt{6} & i\sqrt{2/3}P_{0}k_{z}
& -P_{0}k_{-}/\sqrt{2} & iP_{0}k_{+}/\sqrt{3} & -P_{0}k_{z}/\sqrt{3}\\
&  & P+Q & L-i\sqrt{6}S & M & 0 & iL/\sqrt{2}+\sqrt{3}S & -i\sqrt{2}M\\
&  &  & P-Q & -2\sqrt{2}iS & M & -i\sqrt{2}Q & i\sqrt{3/2}L-S\\
&  &  &  & P-Q & -L-i\sqrt{6}S & -i\sqrt{3/2}L^{+}+S^{+} & -i\sqrt{2}Q\\
&  &  &  &  & P+Q & -i\sqrt{2}M^{+} & -iL^{+}/\sqrt{2}-\sqrt{3}S^{+}\\
&  &  &  &  &  & P & 2\sqrt{2}iS\\
&  &  &  &  &  &  & P
\end{bmatrix}
,
\]
\end{widetext}
\begin{subequations}
\begin{equation}
A_{c}=\hbar ^{2}/(2m_{0})\gamma _{c},  \label{param1}
\end{equation}%
\begin{equation}
P=-\hbar ^{2}/(2m_{0})\mathbf{k}\gamma _{1}\mathbf{k},  \label{param2}
\end{equation}%
\begin{equation}
Q=-\hbar ^{2}/(2m_{0})(\gamma _{2}k_{\parallel }^{2}-2k_{z}\gamma _{2}k_{z}),
\label{param3}
\end{equation}%
\begin{equation}
L=i\sqrt{3}\hbar ^{2}/(2m_{0})(\gamma _{3}k_{z}+k_{z}\gamma _{3})k_{-},
\label{param4}
\end{equation}%
\begin{equation}
M=-\sqrt{3}\hbar ^{2}/(2m_{0})[\gamma
_{2}(k_{x}^{2}-k_{y}^{2})-2ik_{x}\gamma _{3}k_{y}),  \label{param5}
\end{equation}%
\begin{equation}
S=-\hbar ^{2}/(2m_{0})(\kappa k_{z}-k_{z}\kappa )k_{-}.  \label{param6}
\end{equation}%
Here $\gamma _{1},\gamma _{2},\gamma _{3},\kappa $ are modified Luttinger
parameters, $(\gamma _{c}-1)$ describes the remote band contribution to the
CB effective mass, $k_{z}=-i\partial /\partial z$, and $k_{x},k_{y}$ are
c-numbers.

Neglecting the off-diagonal elements in the valence bands (VB's) and
eliminating the VB components of the envelope function, the effective CB
Hamiltonian is obtained as\cite{Darnhofer}
\end{subequations}
\begin{equation}
H_{\text{eff}}(\mathbf{k}_{\parallel })=E_{c}(z)+V(z)+\mathbf{k}\frac{\hbar
^{2}}{2m^{\ast }(z)}\mathbf{k}+\alpha _{0}(z)(\mathbf{k}_{\parallel }\times 
\mathbf{e}_{z})\cdot \mathbf{\sigma },  \label{Heff}
\end{equation}%
where $E_{p}=2m_{0}P_{0}^{2}/\hbar ^{2}$ and $m^{\ast }(z)$ is the effective
mass given by 
\begin{equation*}
\frac{m_{0}}{m^{\ast }(z)}=\gamma _{c}+\frac{2E_{p}}{3U_{\text{lh}}}+\frac{%
E_{p}}{3U_{\text{so}}},
\end{equation*}%
with $U_{\text{lh}}(z)=E-H_{\text{lh}}$, $U_{\text{so}}(z)=E-H_{\text{so}}$, 
$H_{\text{lh}}=E_{v}+V+P-Q,$ and $H_{\text{so}}=E_{v}-\Delta +V+P$. Here $E$
is the eigenenergy and the operators $k_{z}$ in $P$ and $Q$ have been
replaced by $\pi /W$ ($W$ is the well width). The Rashba spin-orbit
interaction strength $\alpha _{0}(z)=\hbar ^{2}/(6m_{0})\partial \gamma
(z)/\partial z,$ where $\gamma (z)=E_{p}[1/U_{\text{lh}}(z)-1/U_{\text{so}%
}(z)]$. It is interesting to notice that RSS at small wave vectors comes
from the coupling to the light hole and spin-orbit split-off VB's, while the
heavy hole bands do not contribute. This is because the basis functions for
the electron $S\uparrow ,$ $S\downarrow $ and those for the heavy hole $%
\left\vert 3/2,\pm 3/2\right\rangle $ are spin eigenstates, while the $%
\mathbf{k}\cdot \mathbf{p}$ interaction between the CB and VB's is
independent of spin. The last term of Eq. (\ref{Heff}) leads to
spin-dependent boundary conditions. To obtain an analytical expression for
RSS, we neglect its influence on the envelope function\cite{Bassani,Pfeffer2} but we keep it in the
effective Hamiltonian (we have checked numerically that this approximation would not change
the qualitative behavior of the resulting analytical expression at large wave vectors),
then the RSS of the n-th subband is given by $\Delta
E_{n}(\mathbf{k}_{\parallel })=\Delta E_{n}^{(1)}(\mathbf{k}_{\parallel
})+\Delta E_{n}^{(2)}(\mathbf{k}_{\parallel })$, where 
\begin{align}
\Delta E_{n}^{(1)}(\mathbf{k}_{\parallel })& =\frac{\hbar ^{2}}{3m_{0}}%
k_{\parallel }\sum\limits_{j}\left\vert F_{n}(z_{j})\right\vert ^{2}[\gamma
(z_{j}^{+})-\gamma (z_{j}^{-})],  \label{RSSa} \\
\Delta E_{n}^{(2)}(\mathbf{k}_{\parallel })& =\frac{\hbar ^{2}}{3m_{0}}%
E_{p}eFk_{\parallel }\int dz\ \left\vert F_{n}(z)\right\vert ^{2}(U_{\text{lh%
}}^{-2}-U_{\text{so}}^{-2}).  \label{RSSb}
\end{align}%
Here $F_{n}(z)$ is the envelope function of the n-th subband along the $z$
axis, $z_{j}^{\pm }=z_{j}\pm 0^{+}$, and $\left\{ z_{j}\right\} $ denote the 
$z$ coordinates of the interfaces. $\Delta E_{n}^{(1)}(\mathbf{k}_{\parallel
})$ can be viewed as the $\Gamma _{8}$ and $\Gamma _{7}$ VBO's-induced
interface electric field contribution, while $\Delta E_{n}^{(2)}(\mathbf{k}%
_{\parallel })$ is roughly proportional to the external electric field. For
small $V(z)$ compared with the band gap, we obtain $\Delta E_{n}^{(2)}(%
\mathbf{k}_{\parallel })$ analytically as 
\begin{equation}
\Delta E_{n}^{(2)}(\mathbf{k}_{\parallel })=\frac{\hbar ^{2}}{3m_{0}}%
E_{p}eFk_{\parallel }\sum\limits_{j}P_{n}^{j}\left[ (U_{\text{lh}%
}^{j})^{-2}-(U_{\text{so}}^{j})^{-2}\right] .  \label{RSSb2}
\end{equation}%
Here $P_{n}^{j}=\int_{\text{layer j}}dz\left\vert F_{n}(z)\right\vert ^{2}$
is the probability of the electron in the $j$-th layer, $U_{\text{lh}}^{j},$ 
$U_{\text{so}}^{j}$ [in which $V(z)$ has been dropped] are for the $j$-th
layer. Further, if $\gamma _{c}$ and the wave function penetration into the
barriers are neglected, and $U_{\text{lh}}^{j}$, $U_{\text{so}}^{j}$ are
replaced by $1/E_{g}$ and $1/(E_{g}+\Delta )$ ($E_{g}$ and $\Delta $ refer
to the well material), respectively, then we recover the result of Ref. 12, 
\begin{equation}
\Delta E_{n}^{(2)}(\mathbf{k}_{\parallel })=\frac{\hbar ^{2}}{m^{\ast }}%
\frac{\Delta }{E_{g}}\frac{2E_{g}+\Delta }{(E_{g}+\Delta )(3E_{g}+2\Delta )}%
eFk_{\parallel },  \label{RSSb3}
\end{equation}%
where $m^{\ast }$ is the CB effective mass of the well. We notice, however,
that a factor of $3/2$ is missing in the definition of $m^{\ast }$ in Ref.
12. Further, Eq. (\ref{RSSb3}) is invalid for narrow-gap semiconductors or
narrow quantum wells, where the subband energy is comparable to the band gap.

From the above discussions, we see that RSS comes from (i) spin-dependent
kinetic and potential energy; (ii) expectation value of the total electric
field (including the external and interface electric fields) in the $\Gamma
_{8}$ and $\Gamma _{7}$ VB's; (iii) variation of band parameters across the
interfaces. The above analytical expressions show that RSS is determined by
the total electric field in the $\Gamma _{8}$ and $\Gamma _{7}$ VB's, in
agreement with Ref. 11. As a result, we see that the Ando argument\cite{Darr}
fails due to an incorrect assumption that RSS is proportional to the total
electric field in the CB. This deepens the previous argument\cite{Bassani} that the failure of the Ando argument
is caused by the spin-dependent boundary conditions or the vanishing barrier penetration.

\begin{figure}[t]
\includegraphics[width=\columnwidth]{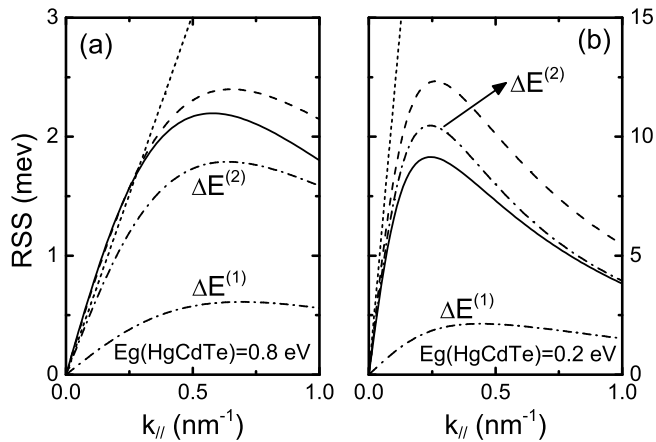}
\caption{RSS of the lowest subband vs. $k_{\parallel }$ for $E_{g}($HgCdTe)=
(a) 0.8 and (b) 0.2 eV. Solid lines denote the full numerical results.
Dashed (short-dashed) lines are obtained from Eqs. (\ref{RSSa}) and
(\ref{RSSb2}) [Eq. (\ref{RSSb3})]. The two contributions $\Delta E^{(1)}$ and
$\Delta E^{(2)}$ are denoted by dash-dotted lines. We take the external
electric field $F$=100 KV/cm, well width $W$=20 nm, VBO=0.3 $\Delta E_{g}$,
where $\Delta E_{g}$=$E_{g}($CdTe$)-E_{g}($HgCdTe$).$}
\end{figure}
To estimate the validity of our analytical expressions, we solve the 8$%
\times $8 Hamiltonian numerically for a biased CdTe/Hg$_{1-\text{x}}$Cd$_{%
\text{x}}$Te/CdTe quantum well and compare the full numerical solutions
with the analytical results in Fig. 1. The band parameters can be found in
Refs. 14 and 20. First, we see that
our analytical expressions [Eqs. (\ref{RSSa}) and (\ref{RSSb2})] agree
better with the numerical results than the previous analytical expression\cite%
{Bassani} [Eq. (\ref{RSSb3})] does at small $k_{\parallel }$. Second,
RSS begins to decrease for $k_{\parallel }$ larger than a critical value $%
k_{0}$. This nonmonotonic behavior is correctly reproduced by our analytical
expression, while the previous analytical expression only gives a linear
behavior. The decrease of RSS at large $k_{\parallel }$ arises as follows. With
increasing $k_{\parallel }$, the subband energy $E_{n}(\mathbf{k}_{\parallel
})\approx E_{n0}+\hbar ^{2}k_{\parallel }^{2}/(2m^{\ast })$ in $U_{\text{lh}}
$ and $U_{\text{so}}$ increases and becomes comparable to $E_{g}($HgCdTe$)$
when $k_{\parallel }\approx k_{0}$. The further increase of $k_{\parallel }$
leads to the decrease of $k_{\parallel }\gamma (z)$ and $k_{\parallel }(U_{%
\text{lh}}^{-2}-U_{\text{so}}^{-2})$ in Eqs. (\ref{RSSa}) and (\ref{RSSb2}).
Consequently, RSS begins to decrease when $k_{\parallel }>k_{0}$.
Further, when $E_{g}$(HgCdTe) decreases from 0.8 to 0.2 eV, the critical
wave vector $k_{0}$ decreases, since a smaller $k_{0}$ is required for $%
E_{n}(\mathbf{k}_{0})$ to become comparable to $E_{g}($HgCdTe$)$. Here we
see that the nonmonotonic behavior of RSS at large $k_{\parallel }$ comes
from the energy dispersion of the subband.
Physically, the coupling to the VB's and, consequently, RSS
are reduced by the increasing energy difference between the CB and VB's.

Third, RSS is dominated by the mean external electric field contribution $%
\Delta E^{(2)}$. This trend becomes more pronounced when $E_{g}($HgCdTe$)$
is decreased from 0.8 to 0.2 eV. It can be understood since the interface
contribution $\Delta E^{(1)}$ is roughly proportional to $1/E_{g}$ through $%
\gamma (z_{j}^{\pm })$ in Eq. (\ref{RSSa}), while the mean external electric
field contribution $\Delta E^{(2)}$ is roughly proportional to $1/E_{g}^{2}$
through $U_{\text{lh}}^{-2}$ and $U_{\text{so}}^{-2}$ in Eq. (\ref{RSSb2}).
Fourth, both $\Delta E^{(1)}$ and $\Delta E^{(2)}$ increase significantly
when $E_{g}($HgCdTe$)$ is decreased from 0.8 to 0.2 eV, in agreement with
Eq. (\ref{RSSb3}). Physically, the increasing RSS is caused by the
increasing coupling between the CB and VB's with decreasing band gap.

In the above, we see that our analytical expression gives a better
description for RSS than the previous analytical expression. Next, by numerically
solving the 8$\times $8 Hamiltonian, we investigate the dependence of RSS on
the band gap $E_{g}$(HgCdTe), VBO, subband index, external electric field,
and well width. The previous analytical expressions\cite{Bassani,Pfeffer2} can
only explain the dependence of RSS on the band gap and external electric
field, while our analytical expression can explain all the dependences, as
we shall show below.

\begin{figure}[t]
\includegraphics[width=\columnwidth]{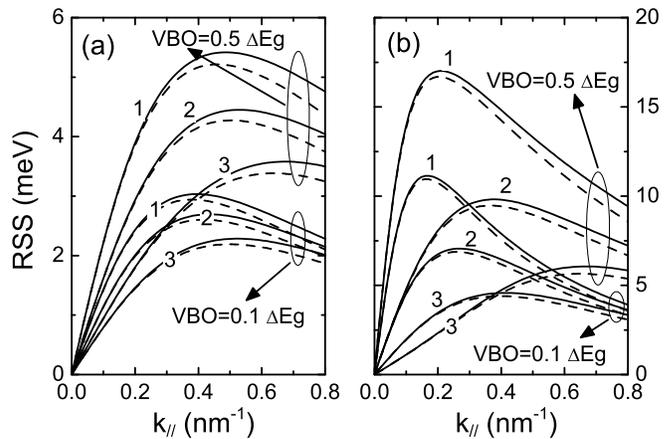}
\caption{RSS of the lowest three subbands (the subband indices are indicated
in the figure) vs. $k_{\parallel }$ along [100] (solid lines) and [110]
(dashed lines) for $E_{g}($HgCdTe)= (a) 0.5 and (b) 0.1 eV and different
VBO's, with fixed well width $W$=20 nm and external electric field $F$=100
KV/cm.}
\end{figure}
In Fig. 2, we plot the RSS of the lowest three subbands for different band
gaps and VBO's. First, in addition to the increase of RSS with
decreasing band gap, it also increases with increasing VBO. This can be
understood from Eqs. (\ref{RSSa}) and (\ref{RSSb2}), because the subband
energy $E_{n}(\mathbf{k}_{\parallel })$ decreases with increasing VBO. As a
result, $\gamma (z_{j}^{\pm })$, $U_{\text{lh}}^{-2}$, and $U_{\text{so}%
}^{-2}$ in Eqs. (\ref{RSSa}) and (\ref{RSSb2}) increases. Second, RSS
decreases with increasing subband index, due to the increase of the subband
energy $E_{n}(\mathbf{k}_{\parallel })$ and the decrease of the asymmetry of
the envelope function at the two interfaces, because the orthogonality
requirement between different eigenstates serves as an effective repulsive
force, which reduces the potential asymmetry produced by the external
electric field.

\begin{figure}[t]
\includegraphics[width=\columnwidth]{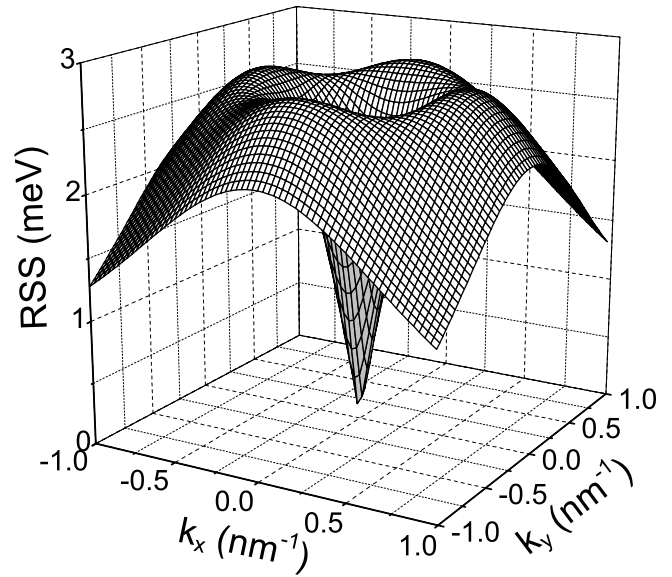}
\caption{RSS for the lowest subband vs. $k_{x}$ and $k_{y}$, with $F$=100
KV/cm, $E_{g}($HgCdTe$)$=$0.5$ eV, and VBO=0.3 $\Delta E_{g}.$}
\end{figure}
Finally, RSS is isotropic at small $k_{\parallel }$ but anisotropic at
large $k_{\parallel }$. This interesting behavior is in contrast to the
current understanding that RSS is always isotropic. From Fig. 3, it can be
seen that RSS has a four-fold anisotropy in the $\mathbf{k}_{\parallel }$
space. This anisotropy comes not from the macroscopic potential, but from
the C$_{\text{4v}}$ symmetry group of the quantum well structure (neglecting
the bulk inversion asymmetry).

\begin{figure}[t]
\includegraphics[width=\columnwidth]{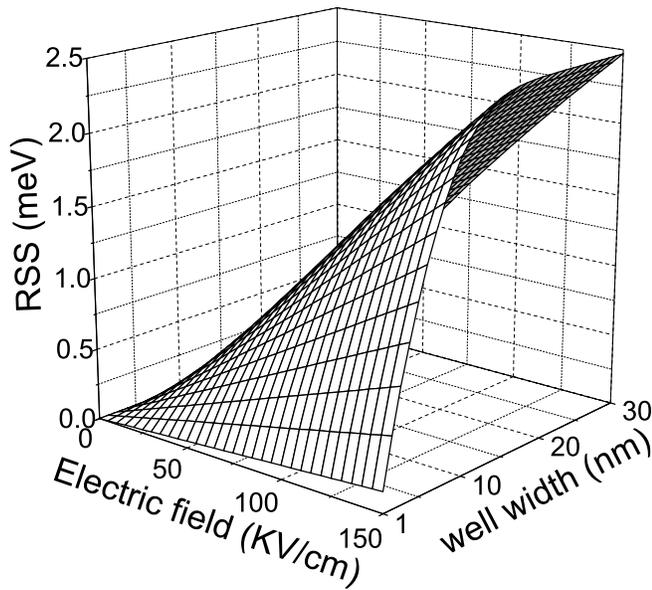}
\caption{RSS for the lowest subband vs. the electric field and well width,
with $k_{\parallel }$=0.1 nm$^{-1}$, $E_{g}($HgCdTe$)$=0.5 eV, VBO=0.3 $%
\Delta E_{g}$.}
\end{figure}
The dependence of the RSS at $k_{\parallel }$=0.1 nm$^{-1}$ on the electric
field and well width is shown in Fig. 4. The RSS increases almost linearly
with increasing electric field, in agreement with Eq. (\ref{RSSb2}), while
it increases with increasing well width and saturates at large well width.
The latter can be explained through the dependence of $\gamma (z)$, $U_{%
\text{lh}}^{-2}$, and $U_{\text{so}}^{-2}$ on the subband energy $E_{n}(%
\mathbf{k}_{\parallel })$. With increasing well width, $E_{n}(\mathbf{k}%
_{\parallel })$ decreases due to the decrease of the confining energy $%
E_{n0} $, such that $\gamma (z),U_{\text{lh}}^{-2}$, $U_{\text{so}}^{-2}$
and, therefore, RSS increase until $E_{n0}$ vanishes and $E_{n}(\mathbf{k%
}_{\parallel })$ approaches a constant value $\hbar ^{2}k_{\parallel
}^{2}/(2m^{\ast })$. Afterwards, $\gamma (z),$ $U_{\text{lh}}^{-2}$, and $U_{%
\text{so}}^{-2}$ do not vary appreciably and RSS saturates. This
behavior is quite different from that of asymmetric AlAs/GaAs/Al$_{0.15}$Ga$%
_{0.85}$As quantum wells, where RSS shows a peak and then decreases with
increasing well width.\cite{Bassani} It was argued that such behavior
comes from the competition between confinement and barrier penetration.
Using our analytical expression, however, the origin of such behavior is
transparent. That is, increasing the well width leads to two competing
effects: the decrease of the subband energy (which increases RSS) and
the asymmetry of the envelope function at the two interfaces (which
decreases RSS).

In summary, based on the full numerical solutions to the eight-band
envelope function model, we have found that at large wave vectors, RSS is both nonmonotonic and anisotropic, in contrast
to the widely used linear and isotropic model. We have derived an analytical expression, which can correctly reproduce such nonmonotonic
behavior at large wave vectors. It shows that the nonmonotonic behavior comes
from the energy dispersion of the subband. We have also investigated numerically the dependence of RSS on the various band parameters and found
that RSS increases with decreasing band gap and subband index, increasing VBO, external electric field, and
well width. Our analytical expression gives better descriptions to all these dependences than the previous analytical expressions.

\end{document}